\documentclass[aps,pra,onecolumn,superscriptaddress]{revtex4}
\usepackage{amsfonts,amssymb,graphicx,amsmath}
\usepackage{epstopdf}
\usepackage{amsmath}
\usepackage[usenames,dvipsnames]{xcolor}
\usepackage{pdfpages}
\usepackage{footmisc}
\usepackage[normalem]{ulem}
\usepackage{amssymb}
\usepackage{color}
\begin{document}

\title{Analytical diagonalization of the open-boundary bosonic Kitaev chain: An asymmetric plane-wave ansatz approach}
\author{Ning Wu}
\email{wunwyz@gmail.com}
\affiliation{Center for Quantum Technology Research, and Key Laboratory of Advanced Optoelectronic Quantum Architecture and Measurements (MOE), School of Physics, Beijing Institute of Technology, Beijing 100081, China}
\author{Wen-Long You}
\email{wlyou@nuaa.edu.cn}
\affiliation{Center for the Cross-disciplinary Research of Space Science and Quantum-technologies (CROSS-Q), College of Physics, Nanjing University of Aeronautics and Astronautics, Nanjing 211106, China}
\begin{abstract}
The bosonic Kitaev chain under open boundary conditions has attracted recent attention due to its realization in driven-dissipative systems and its intriguing non-Hermitian boundary physics. The model is known to be solvable via local squeezing transformations in the position-momentum representation. In this paper, we present an alternative, purely algebraic solution that relies entirely on the standard bosonic Bogoliubov transformation. For an $N$-site chain, we propose an asymmetric plane-wave ansatz with unequal left- and right-moving momenta to analytically solve the associated $2N\times 2N$ non-Hermitian ``associated matrix". The left eigenvalue problem yields $N$ distinct eigenvalues, each of which is twofold degenerate. By carefully resolving these degeneracies using the bosonic commutation relations, we construct the $N$ physical Bogoliubov quasiparticle operators. The construction reveals non-uniquenesses that in special cases exactly mirror the freedom in the local squeezing transformations of the original approach. The diagonal form of the Hamiltonian is obtained explicitly and is shown to be equivalent to the original Hamiltonian. The proposed asymmetric plane-wave ansatz and degeneracy-resolution technique are not limited to the present model and can be generalized to other bosonic pairing systems, including those with inhomogeneous pairing or hopping terms.
\end{abstract}
\maketitle
\section{Introduction}
\par The Kitaev chain, originally proposed as a paradigmatic model for a spinless $p$-wave superconductor, has been instrumental in understanding topological superconductivity and Majorana fermions~\cite{Kitaev}. In recent years, a bosonic analogue of the Kitaev chain, the bosonic Kitaev chain, has attracted significant attention for its realization in driven-dissipative systems and its ability to exhibit non-Hermitian skin effects and chiral transport, despite being governed by a Hermitian Hamiltonian~\cite{PRX2018}. The $N$-site open-boundary bosonic Kitaev chain (OBKC) with purely imaginary hopping and pairing amplitudes,
\begin{eqnarray}\label{Hboson}
H_{B}&=&\frac{1}{2}\sum^{N-1}_{j=1}(it a^\dag_{j+1}a_j+i\Delta a^\dag_{j+1}a^\dag_j-it a^\dag_ja_{j+1}-i\Delta a_ja_{j+1}),
\end{eqnarray}
provides a minimal setup where nontrivial boundary physics emerges from bosonic pairing, where $it$ and $i\Delta$ respectively measure the uniform hopping and pairing strengths between nearest-neighbor sites. Throughout this work we assume $t>0$ and $\Delta>0$~\cite{PRX2018} and define
\begin{eqnarray}\label{delmp}
\delta_{\pm}\equiv \frac{1}{2}(t\pm \Delta).
\end{eqnarray}
We also avoid discussing the special case $t=\Delta$.
\par The system has no lower energy bound because of the presence of negative-energy modes in its diagonal form [see Eq.~(\ref{HBbb}) below]~\cite{PRX2018}. Even though in conventional closed-system equilibrium thermodynamics a Hamiltonian without a lower energy bound is unphysical, the OBKC should be viewed as an effective Hamiltonian that describes a driven-dissipative open system~\cite{PRX2018}, so that the unbounded nature is not problematic. Recently, the associated features of the OBKC have been experimentally observed~\cite{Slim,Nature2024}.
\par The original solution of the OBKC employs a local squeezing transformation in the position-momentum representation~\cite{PRX2018}. We briefly review the above procedures for later comparison. Define the position and momentum operators $x_j=\frac{1}{\sqrt{2}}(a_j+a^\dag_j)$ and $p_j=\frac{1}{\sqrt{2}i}(a_j-a^\dag_j)$, $H_B$ can be rewritten as
\begin{eqnarray}\label{Hbosonxp}
H_{B}&=&\sum^{N-1}_{j=1}(\delta_+ x_j p_{j+1}-\delta_-x_{j+1}p_j).
\end{eqnarray}
The next step is to introduce a local squeezing transformation
\begin{eqnarray}\label{lst}
x_j=\lambda(\delta_+/|\delta_-| )^{j/2}\tilde{x}_j,~~p_j=\lambda^{-1}(\delta_+/|\delta_-| )^{-j/2}\tilde{p}_j,
\end{eqnarray}
where $\lambda$ is an arbitrary $j$-independent real constant. The corresponding annihilation operator is connected with the original ones via
\begin{eqnarray}\label{Ajaj}
\tilde{a}_j=\frac{1}{2}\{[\lambda(\delta_+/|\delta_-| )^{j/2}+\lambda^{-1}(\delta_+/|\delta_-| )^{-j/2}]a_j-[\lambda(\delta_+/|\delta_-| )^{j/2}-\lambda^{-1}(\delta_+/|\delta_-| )^{-j/2}]a^\dag_j \}.
\end{eqnarray}
Under the squeezing transformation, the Hamiltonian is converted to
\begin{eqnarray}
H_B&=& \sum^{N-1}_{j=1}\sqrt{\delta_+|\delta_-|}\left(  \tilde{x}_j  \tilde{p}_{j+1}-\frac{\delta_-}{|\delta_-|} \tilde{x}_{j+1}\tilde{p}_j\right)\nonumber\\
&=& \begin{cases}
\sum^{N-1}_{j=1}\sqrt{\delta_+ \delta_- }i(\tilde{a}^\dag_{j+1}\tilde{a}_j-\tilde{a}^\dag_j \tilde{a}_{j+1}), & \delta_->0,\\
\sum^{N-1}_{j=1}\sqrt{-\delta_+ \delta_- }i(\tilde{a}^\dag_{j+1}\tilde{a}^\dag_j-\tilde{a}_j \tilde{a}_{j+1}), & \delta_-<0.
  \end{cases}
\end{eqnarray}
The second line shows that for $\delta_->0$ ($\delta_-<0$) the model is equivalent to a tight-binding model without the pairing (hopping) term. Physically, the system is said to be ``dynamically unstable" when $\delta_-<0$ since pairs of bosons can be added on two nearest-neighbor sites~\cite{PRX2018}. Actually, Colpa has discussed the diagonalizability of bosonic quadratic forms~\cite{Colpa}, and it turns out that $H_B|_{\delta_-<0}$ ($H_B|_{\delta_->0}$) cannot (can) be written into a diagonal form in terms of quasi-bosons through the Bogoliubov transformation (see below). These facts can also be seen from the $(x,p)$ representation. Consider $N=2$ for simplicity, it is interesting to note that $H_B|_{\delta_->0}$ is proportional to the generator of spatial rotation $x_1p_2-x_2p_1$ (an angular momentum), while  $H_B|_{\delta_-<0}$ is proportional to the generator of Lorentz boost $x_1p_2+x_2p_1$ (by viewing $x_2$ as the time coordinate $t$)~\cite{Ryder}. It is thus obvious that $H_B|_{\delta_-<0}$ has continuous spectrum and cannot be diagonalized (otherwise it should has a discrete spectrum).
\par In the following we focus on the case of $\delta_->0$. $H_B|_{\delta_->0}$ can be diagonalized by further introducing a gauge transformation $\tilde{a}_j=i^j b_j$, so that
\begin{eqnarray}\label{HBbb}
H_B&=& \sum^{N-1}_{j=1}\sqrt{\delta_+ \delta_- }(b^\dag_{j+1}b_j+b^\dag_j b_{j+1})=\sum^N_{n=1} E_n\tilde{b}^\dag_n\tilde{b}_n,
\end{eqnarray}
where $\tilde{b}_n\equiv\sqrt{\frac{2}{N+1}}\sum^N_{j=1}\sin\frac{nj\pi}{N+1}b_j$ and the single-particle dispersion reads $E_n= 2\sqrt{\delta_+ \delta_- } \cos \frac{n\pi}{N+1}$. For even $N$, the $E_n$'s are all nonzero and appear in pairs, $\{E_n,E_{N+1-n}=-E_n|n=1,2,\ldots,N/2\}$. For odd $N$, besides the $(N-1)/2$ pairs $\{E_n,E_{N+1-n}=-E_n|n=1,2,\ldots,(N-1)/2\}$, there is a zero mode $\tilde{b}_{(N+1)/2}$ with $E_{(N+1)/2}=0$. Using Eq.~(\ref{Ajaj}), the quasi-boson annihilation operator $\tilde{b}_n$ can be obtained as
\begin{eqnarray}\label{bnlong}
\tilde{b}_n=\frac{1}{2}\sqrt{\frac{2}{N+1}}\sum^N_{j=1}\sin\frac{nj\pi}{N+1}(-i)^j\left\{\left[\lambda\left(\frac{\delta_+}{ \delta_- } \right)^{j/2}+\frac{1}{\lambda}\left(\frac{ \delta_- }{\delta_+} \right)^{j/2}\right]a_j-\left[\lambda\left(\frac{\delta_+}{ \delta_- } \right)^{j/2}-\frac{1}{\lambda}\left(\frac{ \delta_- }{\delta_+} \right)^{j/2}\right]a^\dag_j \right\}.
\end{eqnarray}
The non-uniqueness of $\tilde{b}_n$ resulting from the arbitrary parameter $\lambda$ implies that the system must have negative single-particle energies in its diagonal form, so that the eigenenergy has no lower bound and the vacuum state is not well defined. On the contrary, the vacuum state or the ground state is well defined if all the $E_n$'s are nonnegative (e.g., for a harmonic oscillator). As mentioned above, the emergence of unbounded spectrum of the OBKC is acceptable since the model serves as an effective Hamiltonian for a driven-dissipative system.
\par While elegant, this approach relies on the $(x,p)$-representation and the geometric intuition of ``squeezing", which, from a purely algebraic standpoint, is somewhat indirect and tricky. The method seems simpler since some elaborate manipulations have been hidden in the squeezing, gauge, and Fourier transformations. Moreover, the arbitrary constant $\lambda$ in the squeezing transformation reflects a non-uniqueness that is not fully discussed in the original solution.
\par Along the other line, Yokomizo and Murakami recently developed a non-Bloch band theory~\cite{PRB2019} and applied it to bosonic Bogoliubov-de Gennes systems~\cite{PRB2021}. By generalizing the Brillouin zone to the complex plane, i.e., the so-called generalized Brillouin zone, they systematically analyzed open-boundary spectra and non-Hermitian skin effects in the thermodynamic limit. For the special case of the OBKC their theory yields the continuous energy bands and indicates the existence of an analytical Bogoliubov transformation. However, the explicit construction of this transformation, the closed-form expressions for the eigenvectors, and the resolution of the degenerate eigenstates were not fully elaborated within their general framework. Furthermore, the non-uniqueness of the Bogoliubov transformation, which is intimately related to the freedom in the squeezing transformation, was not addressed.
\par In this work, we present an alternative, self-contained, and purely algebraic solution of the OBKC that operates entirely within the bosonic Fock space~\cite{Colpa}, without transforming to the $(x,p)$-representation or invoking the generalized Brillouin zone formalism. Our approach consists of the following steps:
\par 1. We cast the Hamiltonian into the standard bosonic quadratic form and perform a conventional Bogoliubov transformation, leading to a non-Hermitian $2N\times 2N$ associated matrix $M$~\cite{Maldonadoa}.
\par 2. To solve the left eigenvalue problem of $M$, $\phi_k M=\Lambda_k\phi_k$, we propose an asymmetric plane-wave ansatz with unequal left- and right-moving momenta, which is a natural generalization of the symmetric plane waves used for Hermitian tridiagonal systems~\cite{Grimm,PRB2024_1,PRB2024_2,physica2025}.
\par 3. Substituting the ansatz into the bulk equations and boundary conditions yields explicit quantization conditions. Solving these conditions explicitly gives the discrete set of allowed momenta and the closed-form eigenvalues.
\par 4. A key technical subtlety emerges: each eigenvalue is twofold degenerate in the left-eigenspace, with two linearly independent eigenvectors. We demonstrate, through a rigorous algebraic proof, that the bosonic commutation relations allow only a single physical Bogoliubov mode to be constructed from the two degenerate eigenvectors. This resolves the degeneracy in a systematic and mathematically transparent manner.
\par 5. The resulting $N$ Bogoliubov transformations contain $N/2$ [$(N-1)/2$] real free parameters for even (odd) $N$. This non-uniqueness is shown to be equivalent to the arbitrary constant $\lambda$ in the squeezing transformation of Ref.~\cite{PRX2018} as a special case, and reflects the fact that the system has no well-defined ground state.
\par Our method distinguishes itself from previous works in several respects. Compared to the squeezing transformation~\cite{PRX2018}, it works directly with creation and annihilation operators, avoiding the intermediate $(x,p)$-representation. Furthermore, there are $\sim N/2$ free parameters in the definition of the quasi-particle operators, while only a single global free parameter $\lambda$ emerges in the squeezing transformation method. Compared to the non-Bloch band theory~\cite{PRB2021}, it provides exact closed-form expressions for eigenvalues and eigenvectors for any finite chain length $N$, rather than the continuous spectrum in the thermodynamic limit. Furthermore, our explicit resolution of the degeneracy and discussion of the non-uniqueness of the Bogoliubov transformation fill technical gaps that were not addressed in earlier studies.
\par The proposed asymmetric plane-wave ansatz and the degeneracy-resolution technique are not limited to the OBKC. In particular, since the symmetric plane-wave ansatz has been successfully applied to the diagonalization of a class of inhomogeneous tridiagonal Hermitian Bloch matrices~\cite{PRB2024_1,PRB2024_2,physica2025}, it is expected that the proposed asymmetric plane-wave ansatz works also for certain inhomogeneous open-boundary bosonic pairing models with nonuniform on-site energies or hopping/pairing strengths near the chain ends, where inhomogeneous non-Hermitian associated matrices are relevant. We demonstrate the applicability of our method in a minimal inhomogeneous OBKC with nonuniform hopping and pairing on the first bond of the lattice.
\par The rest of the paper is organized as follows. In Sec.~\ref{SecII}, we review the bosonic quadratic form and set up the associated matrix $M$ to be solved. In Sec.~\ref{SecIII}, we solve the left eigenvalue problem of $M$ using the asymmetric plane-wave ansatz, derive the quantization conditions, construct the physical Bogoliubov modes, and discuss the non-uniqueness of the Bogoliubov transformations and the correspondence with the squeezing transformation. Sec.~\ref{SecIV} concludes the paper and discusses the inhomogeneous case briefly.
\section{Bosonic quadratic form and Bogoliubov transformation}\label{SecII}
\par Consider a general $N$-mode bosonic quadratic form
\begin{eqnarray}\label{Hform}
H=\sum^N_{i,j=1}\left[a^\dag_i A_{ij}a_j+\frac{1}{2}(a^\dag_i B_{ij}a^\dag_j+a_j B^*_{ij}a_i)\right],
\end{eqnarray}
where $a_i$ and $a^\dag_i$ are respectively the annihilation and creation operators of the $i$th mode and satisfy the usual commutation relations $[a_i,a_j]=0$ and $[a_i,a^\dag_j]=\delta_{ij}$. To ensure the Hermiticity of $H$, the two square matrices $A$ and $B$ should satisfy
\begin{eqnarray}
A_{ij}=A^*_{ji},~B_{ij}=B_{ji},
\end{eqnarray}
i.e., $A$ is Hermitian and $B$ is symmetric.
\par Consider the following Bogoliubov transformation,
\begin{eqnarray}\label{bogov}
\eta_k&=&\sum_{i}(g_{ki}a_i+h_{ki}a^\dag_i),~~\eta^\dag_k=\sum_{i}(g^*_{ki}a^\dag_i+h^*_{ki}a_i),
\end{eqnarray}
where $k=1,2,\ldots,N$ and the undetermined coefficients $g_{ki}$'s and $h_{ki}$'s are generally complex. To simplify the notations, we define two $1\times N$ row vectors $g_k \equiv(g_{k1},\ldots,g_{kN})$ and $h_k \equiv(h_{k1},\ldots,h_{kN})$. We require that the $\eta$-particles are also bosons, i.e., $[\eta_k,\eta^\dag_{k'}]=\delta_{kk'}$ and $[\eta_k,\eta_{k'}]=0$, leading to
\begin{eqnarray}
g_kg^\dag_{k'}-h_kh^\dag_{k'}&=&\delta_{kk'},\label{ghcond}\\
g_kh^\mathrm{T}_{k'}-h_kg^\mathrm{T}_{k'} &=&0,\label{ghcondT}
\end{eqnarray}
where the superscript T denotes matrix transposition.
\par In this work, we always assume that after the Bogoliubov transformation the Hamiltonian admits a diagonal form,
\begin{eqnarray}\label{HLambda}
H=\sum^N_{k=1}\Lambda_k \eta^\dag_k \eta_k+\mathrm{const.},
\end{eqnarray}
where the single-particle energy $\Lambda_k$ must be real due to the Hermiticity of $H$. This requirement is not trivial since not all Hermitian bosonic quadratic forms can be diagonalized (recall the case of $\delta_-<0$ discussed above)~\cite{Colpa}. Another simpler example is an inverted harmonic oscillator with the reduced Hamiltonian $H_{\mathrm{IHO}}=i[(a^\dag)^2-a^2]$~\cite{IHO}, which cannot be written in the diagonal form (\ref{HLambda}) since its energy spectrum is continuous. However, we allow for negative values of $\Lambda_k$, for which the system is energetically unstable but may be ``dynamically stable"~\cite{Ueda}.
\par The usual way to determine the coefficients is to use the relation~\cite{Lieb}
\begin{eqnarray}\label{Heta}
~[H,\eta_k]= -\Lambda_k\eta_k,
\end{eqnarray}
which gives
\begin{eqnarray}
&&\sum_i\left[\sum_j(A_{ij}h_{kj}-B_{ij}g_{kj})a^\dag_i+\sum_j(B^*_{ij}h_{kj}-A_{ji}g_{kj})a_i\right]=-\Lambda_k\sum_{i}(g_{ki}a_i+h_{ki}a^\dag_i).
\end{eqnarray}
\par By comparing the coefficients of $a_i$ and $a^\dag_i$ on both sides, we obtain
\begin{eqnarray}\label{Lambdagh}
\Lambda_kg_{ki}&=&\sum_j(g_{kj} A_{ji}-h_{kj}B^*_{ji}),~~\Lambda_kh_{ki}= \sum_j(g_{kj}B_{ji}-h_{kj}A^*_{ji}),
\end{eqnarray}
which are actually of the same form as in the fermion case~\cite{Lieb}.
\par We now combine the two vectors $g_k$ and $h_k$ into a single $1\times 2N$ row vector $\phi_k\equiv (g_k,h_k)$, then Eq.~(\ref{Lambdagh}) can be recast in a matrix form
\begin{eqnarray}\label{phikM}
\phi_k\Lambda_k=\phi_k M,
\end{eqnarray}
where $M$ is a $2N\times 2N$ non-Hermitian matrix,
\begin{eqnarray}
M=\left(
                                                                                                                     \begin{array}{cc}
                                                                                                                       A & B \\
                                                                                                                       -B^* & -A^* \\
                                                                                                                     \end{array}
                                                                                                                   \right),
\end{eqnarray}
which is sometimes called the associated matrix~\cite{Maldonadoa}.
\par Several remarks are in order. Note that Eq.~(\ref{Heta}) is only a necessary condition for the Hamiltonian being in the diagonal form (\ref{HLambda}). For example, if extra terms, say $\sum_{kk'}C_{kk'}\eta_k\eta_{k'}$, were added to $H$, the resulting condition is still given by Eq.~(\ref{Heta}). In addition, since the dimension of the associated matrix has been doubled to $2N$ (the number physical modes is always $N$), there exist redundant degrees of freedom in the determination of the physical eigenstates using the Bogoliubov transformation. These issues will be carefully addressed within our approach.

\par Now, defining the unitary matrix $P=P^{-1}=\left(
                                                                                                                                                \begin{array}{cc}
                                                                                                                                                  0 & 1_{N\times N} \\
                                                                                                                                                  1_{N\times N} & 0 \\
                                                                                                                                                \end{array}
                                                                                                                                              \right)
$ that satisfies $PMP^{-1}=-M^*$ and taking the complex conjugate of Eq.~(\ref{phikM}), we have $\phi^*_k\Lambda^*_k=\phi^*_k M^*=-\phi^*_k PMP^{-1}$, giving
\begin{eqnarray}\label{phikMconj}
(\phi^*_kP)(-\Lambda^*_k) =(\phi^*_kP) M.
\end{eqnarray}
The above equation shows that $-\Lambda^*_k$ is also a left eigenvalue of $M$. Since we are seeking real eigenvalues, these eigenvalues will appear in pairs as $(\Lambda_k,-\Lambda_k)$. There should be $N$ such pairs. Note, however, that the vector $\phi^*_kP=(h^*_k,g^*_k)$ does not satisfy Eq.~(\ref{ghcond}), and hence cannot be a physical solution unless the eigenvalue $-\Lambda_k$ is degenerate. As we will see, it is actually the case for the OBKC.
\section{Solution of the open-boundary bosonic Kitaev chain: A plane-wave ansatz}\label{SecIII}
\subsection{The associated matrix $M$}
\par We write the OBKC (\ref{Hboson}) in the form of Eq.~(\ref{Hform}),
\begin{eqnarray}\label{Hform1}
H_{B}&=& \sum_{ij}\frac{it}{2}(\delta_{i-j,1}-\delta_{j-i,1})a^\dag_i a_j+\frac{1}{2}\sum_{ij}\frac{i\Delta}{2}(\delta_{i-j,1}+\delta_{j-i,1})( a^\dag_i a^\dag_j - a_j a_i),
\end{eqnarray}
which gives
\begin{eqnarray}\label{AandB}
A_{ij}&=&\frac{it}{2}(\delta_{i-j,1}-\delta_{j-i,1}),~~B_{ij}= \frac{i\Delta}{2}(\delta_{i-j,1}+\delta_{j-i,1}),
\end{eqnarray}
or, explicitly,
\begin{eqnarray}
A&=& \frac{it}{2}\left(
     \begin{array}{ccccccc}
       0 & -1 &   &   &   &   &  \\
       1 & 0 & -1 &   &   &   &   \\
         & 1 & 0 & -1  &   &   &   \\
         &   &   & \ddots  &   &   &   \\
         &   &   &   &    &   &   \\
         &   &   &   & 1 & 0 & -1 \\
         &   &   &   &   & 1 & 0 \\
     \end{array}
   \right),~~B=\frac{i\Delta}{2}\left(
     \begin{array}{ccccccc}
       0 & 1 &   &   &   &   &   \\
1 & 0 & 1 &   &   &   &   \\
         &1 & 0 & 1  &   &   &   \\
         &   &   & \ddots  &   &   &   \\
         &   &   &   &   &   &   \\
         &   &   &   &1 & 0 & 1 \\
         &   &   &   &   &  1 & 0 \\
     \end{array}
   \right).
\end{eqnarray}
Since $A$ and $B$ are both purely imaginary, we have $M=\left(
                                                                                                                     \begin{array}{cc}
                                                                                                                       A & B  \\
                                                                                                                       B & A \\
                                                                                                                     \end{array}
                                                                                                                   \right)=-M^*$, giving
\begin{eqnarray}\label{phikMconj2}
(\phi^*_k)(-\Lambda^*_k) =(\phi^*_k) M.
\end{eqnarray}
The vector $\phi^*_k=(g^*_k,h^*_k)$ is thus degenerate with $\phi^*_kP$ and it satisfies Eq.~(\ref{ghcond}), so $\phi^*_k$ may be a physical solution.
                                                                                                                                                                                                                                      \par Let                                                                                                                     $Q=\frac{1}{\sqrt{2}}\left(
                                                                                                                     \begin{array}{cc}
                                                                                                                       1_{N\times N} & 1_{N\times N}  \\
                                                                                                                       1_{N\times N} & -1_{N\times N} \\
                                                                                                                     \end{array}
                                                                                                                    \right)$ be a real and unitary matrix, then
\begin{eqnarray}
\tilde{M}\equiv QMQ^{-1}=\left(
                                                                                                                     \begin{array}{cc}
                                                                                                                       F_1 & 0  \\
                                                                                                                       0 & F_2 \\
                                                                                                                     \end{array}
                                                                                                                   \right),
\end{eqnarray}
where $F_1\equiv A+B$, $F_2\equiv A-B$. It is easy to see that $\phi_k Q^{-1}$  is a left eigenvector of $\tilde{M}$ with eigenvalue $\Lambda_k$,
\begin{eqnarray}\label{vQ}
(\phi_kQ^{-1})\tilde{M}=(\phi_kQ^{-1})\Lambda_k.
\end{eqnarray}
Thus, we only need to solve the two $N\times N$ matrices $F_1$ and $F_2$ separately. Note that $F^*_i=-F_i$, so the eigenvalues of $F_i$ also appear in pairs if they are real.
\par $F_1$ and $F_2$ can be written in terms of $\delta_{\pm}$ as
\begin{eqnarray}
F_1&=&i\left(
     \begin{array}{ccccccc}
       0 & -\delta_- &   &   &   &   &  \\
       \delta_+ & 0 & -\delta_- &   &   &   &   \\
         & \delta_+ & 0 & -\delta_-  &   &   &   \\
         &   &   & \ddots  &   &   &   \\
         &   &   &   &    &   &   \\
         &   &   &   & \delta_+ & 0 & -\delta_- \\
         &   &   &   &   & \delta_+ & 0 \\
     \end{array}
   \right),~~~F_2=i\left(
     \begin{array}{ccccccc}
       0 & -\delta_+ &   &   &   &   &  \\
       \delta_- & 0 &-\delta_+ &   &   &   &   \\
         & \delta_- & 0 & -\delta_+  &   &   &   \\
         &   &   & \ddots  &   &   &   \\
         &   &   &   &    &   &   \\
         &   &   &   & \delta_- & 0 & -\delta_+\\
         &   &   &   &   & \delta_- & 0 \\
     \end{array}
   \right).
\end{eqnarray}
It is interesting to note that $F_1$ or $F_2$ describes a single-particle problem in a tight-binding Hatano-Nelson model with purely imaginary asymmetry hopping strengths~\cite{HN}.
\par At this stage it is also useful to observe that for $\delta_->0$ both $F_1$ and $F_2$ are similar to the same Hermitian tridiagonal matrix,
\begin{eqnarray}
S^{-1}F_1S=SF_2S^{-1}=i\sqrt{\delta_+\delta_-}\left(
     \begin{array}{ccccccc}
       0 & -1 &   &   &   &   &  \\
       1 & 0 &-1 &   &   &   &   \\
         & 1 & 0 & -1  &   &   &   \\
         &   &   & \ddots  &   &   &   \\
         &   &   &   &    &   &   \\
         &   &   &   & 1 & 0 & -1\\
         &   &   &   &   & 1 & 0 \\
     \end{array}
   \right),
\end{eqnarray}
where
\begin{eqnarray}
S=\mathrm{diag}(\sqrt{\delta_+/\delta_-},(\sqrt{\delta_+/\delta_-})^2,\ldots,(\sqrt{\delta_+/\delta_-})^N).
\end{eqnarray}
The above similar transformation exactly reflects the local squeezing transformation (\ref{lst}). Nevertheless, we will not take advantage of this similarity in this work. The reason is twofold: on one hand, we want to better clarify the applicability of our method in general situations; on the other hand, such kind of similar transformation may not be easily found in general cases. Therefore, in the following we choose to directly solve the eigenvalue problem of $F_1$ and $F_2$ by using the asymmetric plane-wave ansatz.
\subsection{Solutions of $F_1$ and $F_2$}
\par Consider the left eigenvalue problem of $F_1$,
\begin{eqnarray}\label{bdyeq}
VF_1=V\Lambda,~\Lambda\in \mathbb{R}.
\end{eqnarray}
Explicitly, we have $N-2$ bulk equations,
\begin{eqnarray}\label{bulk}
V_j\Lambda=i(-\delta_-V_{j-1}+\delta_+V_{j+1}),~j=2,\ldots,N-1
\end{eqnarray}
and two boundary equations,
\begin{eqnarray}\label{bdyeq}
V_1\Lambda&=&i\delta_+ V_2,\nonumber\\
V_N\Lambda&=&-i\delta_- V_{N-1}.
\end{eqnarray}
We first note that $\Lambda$ cannot be zero for even $N$. Suppose otherwise $\Lambda=0$, then from Eq.~(\ref{bdyeq}) we have $V_2=V_{N-1}=0$. But Eq.~(\ref{bulk}) tells us that $V_2=V_4=\cdots=V_{N}=0$ and $V_{N-1}=V_{N-3}=\cdots=V_1=0$, we thus get a null vector. For odd $N$, it is easy to check that
\begin{eqnarray}\label{zeromode}
V^{(\Lambda=0)}=\sqrt{\frac{1-(\delta_-/\delta_+)^{2}}{1-(\delta_-/\delta_+)^{N+1}}}\left(1,0,\delta_-/\delta_+,0,(\delta_-/\delta_+)^2,\ldots,0,(\delta_-/\delta_+)^{(N-1)/2}\right)
\end{eqnarray}
is a unique normalized left eigenvector of $F_1$ belonging to the zero eigenvalue $\Lambda=0$. As a result, we focus on solutions with $\Lambda\neq 0$ in the following analysis for both even and odd $N$.
\par When solving Hermitian tridiagonal matrices, a plane-wave ansatz $V_j=Xe^{ipj}+Ye^{-ipj}$ is usually used as a trial eigenvector~\cite{Grimm,PRB2024_1,PRB2024_2,physica2025}, where $X$ and $Y$ are $j$-independent parameters and $p$ is a pseudo-momentum to be determined. The reason why the symmetric plane-wave ansatz works for Hermitian matrices is that when inserting the ansatz into the bulk equations either the right moving branch $e^{ipj}$ or the left moving branch $e^{-ipj}$ yields the same expression for the eigenvalue $\Lambda$~\cite{PRB2024_1,PRB2024_2,physica2025}. This is evidently not the case for the non-Hermitian matrix $F_1$ where the two branches give different forms of $\Lambda$, leading to an absurd condition $e^{ip}=e^{-ip}$. Nevertheless, it turns out that a simple generalization of the ansatz to an asymmetric one solves the problem,
\begin{eqnarray}\label{PLA}
V_j=Xe^{ipj}+Ye^{-ip'j},
\end{eqnarray}
where the complex right- and left-moving pseudo-momenta $p$ and $p'$ are not necessarily equal. We emphasize that even for a symmetric plane-wave ansatz, complex pseudo-momenta of the form $p=m\pi + \tilde{p}~(m\in\mathbb{Z},~\tilde{p}\in\mathbb{R})$ may occur in the solution of certain Hermitian tridiagonal matrices, giving bound states with decaying wave functions~\cite{PRB2024_1,PRB2024_2,physica2025}.
\par Inserting the ansatz into the bulk equations (\ref{bulk}) and comparing the coefficients of $e^{ip j}$ and $e^{-ip'j}$ on both sides, we get two different expressions for $\Lambda$,
\begin{eqnarray}\label{LLambda}
\Lambda&=&i(-\delta_- e^{-ip}+\delta_+e^{ip}),\nonumber\\
\Lambda&=&i(-\delta_- e^{ip'}+\delta_+e^{-ip'}).
\end{eqnarray}
To determine $p$ and $p'$, we further apply the ansatz in the two boundary equations, yielding
\begin{eqnarray}\label{XYeq}
&&\left(
  \begin{array}{cc}
   -i\delta_- & -i\delta_- \\
   i\delta_+e^{ip(N+1)} & i\delta_+e^{-ip'(N+1)}\\
  \end{array}
\right)\left(
         \begin{array}{c}
           X \\
           Y \\
         \end{array}
       \right)=0,
\end{eqnarray}
where we have used Eq.~(\ref{LLambda}) to eliminate $\Lambda$. The above equation has a nontrivial solution with $X=-Y$ when $e^{ip(N+1)}=e^{-ip'(N+1)}$, giving $p+p'=\frac{2\pi m}{N+1},~m\in \mathbb{Z}$. In other words, the imaginary parts of $p$ and $p'$ are opposites of each other, while the sum of their real parts must be an integer multiple of $\frac{2\pi }{N+1}$. If we write $p\equiv\alpha+i\beta$ and $p'\equiv \alpha'+i\beta'$, where $\alpha,\beta,\alpha'$, and $\beta'$ are all real, then
\begin{eqnarray}
\alpha+\alpha'&=&\frac{2\pi m}{N+1},~m\in \mathbb{Z},\label{alphabeta}\\
\beta&=&-\beta'.\label{alphabeta2}
\end{eqnarray}
In the following, we let $\alpha$ and $\alpha'$ both lie in the first Brillouin zone $[-\pi,\pi)$.
\par Since $\Lambda$ is required to be real, the imaginary parts of the two expressions in Eq.~(\ref{LLambda}) vanish,
\begin{eqnarray}\label{cosaa}
 \cos\alpha(-\delta_- e^{\beta }+\delta_+  e^{-\beta })&=&0,\nonumber\\
\cos\alpha'( -\delta_- e^{ \beta }+\delta_+  e^{-\beta })&=&0,
\end{eqnarray}
where we used Eq.~(\ref{alphabeta2}). The real parts read
\begin{eqnarray}
\Lambda&=&-\sin\alpha(\delta_- e^{\beta }+\delta_+  e^{-\beta }),\nonumber\\
\Lambda&=&\sin\alpha'(\delta_- e^{\beta }+\delta_+  e^{-\beta }).
\end{eqnarray}
Recall that $\Lambda\neq0$, we have $\sin\alpha=-\sin\alpha'$. It is possible that $\alpha=-\alpha'$ or $\alpha=\alpha'\pm \pi$, where the sign is chosen such that $\alpha,\alpha'\in[-\pi,\pi)$. Suppose the former is true, then $\beta=-\beta'$ means that $p=-p'$, so that $V_j=(X+Y)e^{ipj}=0$ for any $j$. Thus, the only possibility is
\begin{eqnarray}\label{alpha1alpha2}
\alpha=\alpha'\pm \pi.
\end{eqnarray}
\par It is easy to see that $\alpha \neq \pm \pi/2$. Suppose otherwise $\alpha = \pm \pi/2$, the above equation gives $\alpha'=\mp\pi/2=-\alpha$, leading again to $p=-p'$ and hence a trivial solution $V_j=0$. As a result, all the physically allowed solutions must satisfy $\cos\alpha\neq0$ and $\cos\alpha'\neq0$. According to Eq.~(\ref{cosaa}), we have $\delta_-e^{\beta}=\delta_+e^{-\beta}$, or
\begin{eqnarray}\label{expB}
\frac{\delta_+}{\delta_-}=e^{2\beta}.
\end{eqnarray}
This important relation not only uniquely determines the real parameter $\beta$ but also implies that the OBKC cannot be diagonalized via the Bogoliubov transformation for $\delta_-<0$, consistent with the foregoing analysis based on the squeezing transformation~\cite{PRX2018}. From Eq.~(\ref{expB}) we have $ \delta_+e^{-\beta}=\delta_-e^{\beta}=\sqrt{\delta_+\delta_-}$, so that
\begin{eqnarray}\label{Lambdaalpha}
\Lambda(\alpha)&=&-2\sin\alpha \sqrt{\delta_+\delta_-}.
\end{eqnarray}
Using the relation $e^{-i\alpha' j}=e^{-i(\alpha\pm \pi) j}=(-1)^j e^{-i\alpha j}$, we get the wave function expressed in terms of $\alpha$ and $\beta$:
\begin{eqnarray}\label{VjX}
V_j(\alpha)=[e^{i\alpha j}-(-1)^je^{-i\alpha j}]e^{-\beta j},
\end{eqnarray}
where the overall factor $X=-Y$ has been set to unity. Although the above eigenvector is obtained from the plane-wave ansatz under the assumption of $\Lambda\neq0$, it is interesting to note that the zero mode $V^{(\Lambda=0)}$ given by Eq.~(\ref{zeromode}) can be incorporated into Eq.~(\ref{VjX}) by setting $\alpha=0$.
\par It is easy to check that
\begin{eqnarray}
\Lambda(-\alpha)&=&-\Lambda(\alpha),\label{LLstar}\\
V_j(-\alpha)&=&V^*_j(\alpha),\label{VVstar}\\
\Lambda(\pi-\alpha)&=&\Lambda(\alpha),\label{LpiL}\\
V_j(\pi-\alpha)&=&-V_j(\alpha)\label{VpiV}.
\end{eqnarray}
The last two relations indicate that $\alpha$ and $\pi-\alpha$ give the same eigenvalue and eigenvector, so that we can restrict $\alpha$ to the interval $(-\frac{\pi}{2},\frac{\pi}{2})$. By combining Eq.~(\ref{alphabeta}) with (\ref{alpha1alpha2}), we finally obtain the $N$ allowed values of $\alpha$ (for either even or odd $N$, see Fig.~1),
\begin{eqnarray}\label{alpham}
\alpha_m=-\frac{\pi}{2}+\frac{m\pi}{N+1},~m=1,2,\ldots,N.
\end{eqnarray}
\begin{figure}\label{alphaplot1}
\includegraphics[width=.95\textwidth]{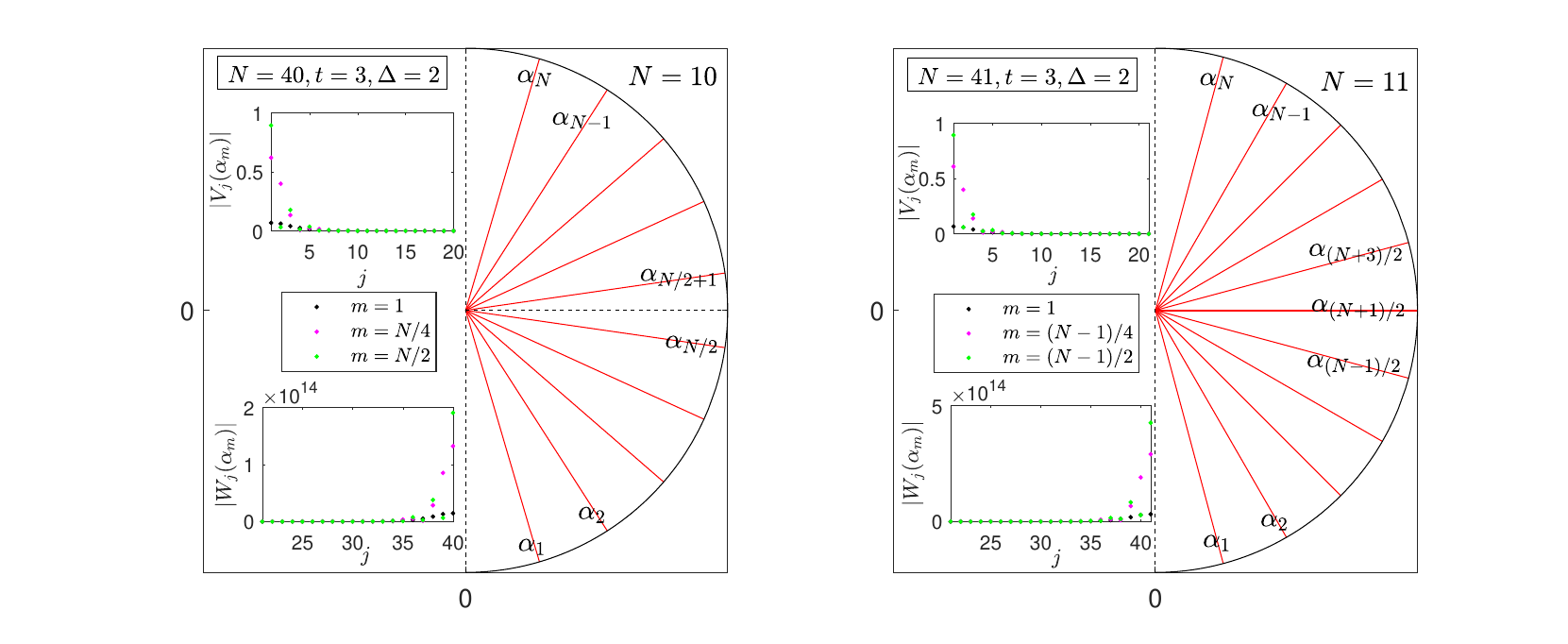}
\caption{Allowed values of $\alpha$ in the plane-wave ansatz. (a) $N=10$, (b) $N=11$. For even $N$, all the $\alpha_m$'s are nonzero; while for odd $N$ there exists a zero mode with $\alpha_{(N+1)/2}=0$. The insets show examples of the absolute values of $V_j(\alpha_m)$ and $W_j(\alpha_m)$, which are localized near the left and right ends of the chain, respectively.}
\label{Fig1}
\end{figure}
\par In summary, $F_1$ does have $N$ real left eigenvalues $\Lambda(\alpha)=-2\sin\alpha \sqrt{\delta_+\delta_-}$, with $\alpha$ being given by Eq.~(\ref{alpham}). For even $N$, $\Lambda(\alpha_i)$ is positive (negative) for $i=1,2,\ldots,\frac{N}{2}$ ($i=\frac{N}{2}+1,\frac{N}{2}+2,\ldots,N$) and we can group them into $N/2$  pairs $\{\Lambda(\alpha_n),\Lambda(\alpha_{N+1-n})=-\Lambda(\alpha_n)|n=1,2,\ldots,\frac{N}{2}\}$. Similarly, for odd $N$ the $N-1$ nonzero eigenvalues can be grouped into $(N-1)/2$ pairs $\{\Lambda(\alpha_n),\Lambda(\alpha_{N+1-n})=-\Lambda(\alpha_n)|n=1,2,\ldots,\frac{N-1}{2}\}$.
\par Inserting $\alpha_m$ into Eq.~(\ref{VjX}), we obtain
\begin{eqnarray}\label{VjXm}
V_j(\alpha_m)=2 (-i)^{j-1}\sin\frac{\pi mj}{N+1} e^{-\beta j}.
\end{eqnarray}
\par The eigenvalue problem of $F_2$ can be solved in a similar way by switching the roles of $\delta_+$ and $\delta_-$. The eigenvalues of $F_2$ are still given by $\Lambda(\alpha_m)$ and the corresponding eigenvector reads
\begin{eqnarray}\label{WjXm}
W_j(\alpha_m)=2 (-i)^{j-1}\sin\frac{\pi mj}{N+1} e^{\beta j}.
\end{eqnarray}
It is obvious that $V$ ($W$) is localized near the left (right) edge of the chain (insets of Fig.~1).
\par It is useful to make explicit the connection with the non-Bloch band theory~\cite{PRB2021} in the thermodynamic limit $N\to \infty$, where the $\alpha$ becomes continuous on $(-\frac{\pi}{2},\frac{\pi}{2})$. The spectrum $\Lambda(\alpha)=-2\sin\alpha \sqrt{\delta_+\delta_-}$ with $\alpha\in(-\frac{\pi}{2},\frac{\pi}{2})$ is precisely the continuous open-boundary band obtained from the finite-$N$ levels. In the non-Bloch band theory, the generalized Brillouin zone for the OBKC consists of two circles with radii $|z_+|=\sqrt{\delta_+/\delta_-}=e^{\beta}$ and $|z_-|=\sqrt{\delta_-/\delta_+}=e^{-\beta}$ in the two bosonic Bogoliubov-de Gennes sectors~\cite{PRB2021}. These are exactly the skin factors appearing in the two wave functions $V$ and $W$. Thus, the real part of the pseudo momentum in the ansatz corresponds to the continuum momentum in the non-Bloch band theory, while the skin factors fix the non-Bloch radial deformation of the two sectors.
\subsection{Construction of $g_k$ and $h_k$}
\par Using Eq.~(\ref{vQ}), we obtain the $2N$ left (unnormalized) eigenvectors of $M$,
\begin{eqnarray}\label{twoNrow}
\left(
              \begin{array}{c}
                \chi(\alpha_1) \\
                \vdots \\
                \chi(\alpha_N) \\
                \xi(\alpha_1) \\
                \vdots \\
                \xi(\alpha_N) \\
              \end{array}
            \right)M=\left(
              \begin{array}{c}
               \Lambda(\alpha_1) \chi(\alpha_1) \\
                \vdots \\
                \Lambda(\alpha_N) \chi(\alpha_N) \\
               \Lambda(\alpha_1) \xi(\alpha_1) \\
                \vdots \\
              \Lambda(\alpha_N)  \xi(\alpha_N) \\
              \end{array}
            \right),
\end{eqnarray}
where
\begin{eqnarray}\label{chixi}
\chi(\alpha_m)&\equiv&(V(\alpha_m), V(\alpha_m)),\nonumber\\
\xi(\alpha_m)&\equiv&(W(\alpha_m), -W(\alpha_m)),
\end{eqnarray}
are two $1\times 2N$ degenerate eigenvectors belonging to the common eigenvalue $\Lambda(\alpha_m)$. It is obvious that $\chi(\alpha_m)\xi^\dag(\alpha_m)=\chi(\alpha_m)\xi^{\mathrm{T}}(\alpha_m)=0$.
\par To proceed, we need two useful identities (see Appendix~\ref{AppA} for the proof):
\begin{eqnarray}
V(\alpha_m)W^\dag(\alpha_n)&=&W(\alpha_m)V^\dag(\alpha_n)=2(N+1)\delta_{mn}, \label{VW1}\\
V(\alpha_m)W^{\mathrm{T}}(\alpha_n)&=&W(\alpha_m)V^{\mathrm{T}}(\alpha_n)=2(N+1)\delta_{m+n,N+1}.\label{VW2}
\end{eqnarray}
\par For an $N$-mode model, there should be $N$ Bogoliubov quasiparticle operators $\eta_k~(k=1,2,\ldots,N)$ in the diagonal form Eq.~(\ref{HLambda}). Obviously, the $N$ sets of coefficients $\phi_k=(g_k,h_k)$ in $\eta_k=\sum_{i}(g_{ki}a_i+h_{ki}a^\dag_i)$ must be constructed from the $2N$ row vectors in Eq.~(\ref{twoNrow}). Since the two mathematically acceptable degenerate left eigenvectors $\chi(\alpha_m)$ and $\xi(\alpha_m)$ are orthogonal, one may naively think that for a fixed $m$ two physically allowed linear combinations of them can be constructed,
\begin{eqnarray}\label{gggpgp}
(g_m,h_m)&=&c_{m,1}\chi(\alpha_m)+c_{m,2}\xi(\alpha_m)=(c_{m,1}V(\alpha_m)+c_{m,2}W(\alpha_m),c_{m,1}V(\alpha_m)-c_{m,2}W(\alpha_m)),\nonumber\\
(g'_m,h'_m)&=&c'_{m,1}\chi(\alpha_m)+c'_{m,2}\xi(\alpha_m)=(c'_{m,1}V(\alpha_m)+c'_{m,2}W(\alpha_m),c'_{m,1}V(\alpha_m)-c'_{m,2}W(\alpha_m)),
\end{eqnarray}
where we are free to choose $c_{m,1}$ and $c'_{m,1}$ to be real since multiplying $\eta_k$ by an overall phase factor does not change the bosonic commutation relations. However, the foregoing argument indicates that we cannot achieve this; otherwise, the allowed number of $\eta_k$'s will be $2N$. We prove this by the method of contradiction. Suppose that the two sets of coefficients in Eq.~(\ref{gggpgp}) are both physically allowed, then the requirement (\ref{ghcond}) tells us that
\begin{eqnarray}\label{gghh6}
g_m g_m^\dag -h_m h_m^\dag &=&1,\nonumber\\
g'_m g'^\dag_m -h'_m h'^\dag_m &=&1,\nonumber\\
g_m g'^\dag_m-h_mh'^\dag_m&=&0,
\end{eqnarray}
which after using the relation (\ref{VW1}) give
\begin{eqnarray}
c_{m,1}(c_{m,2}+c^*_{m,2})&=&\frac{1}{4(N+1)},\label{c1c2cond1}\\
c'_{m,1}(c'_{m,2}+c'^*_{m,2})&=&\frac{1}{4(N+1)},\label{c1c2cond2}\\
c_{m,1}c'^*_{m,2}+c'_{m,1} c_{m,2}&=&0.\label{c1c2cond3}
\end{eqnarray}
\par We now show that the above three equations have no solutions, and hence we can only construct a single physical eigenvector out of the two degenerate vectors $\chi(\alpha_m)$ and $\xi(\alpha_m)$. First note that $c_{m,1}\neq0$ and $c'_{m,1}\neq0$, then from Eq.~(\ref{c1c2cond3}) we have $c'^*_{m,2}=-c_{m,2}c'_{m,1}/c_{m,1}$ and $c'_{m,2}=-c^*_{m,2}c'_{m,1}/c_{m,1}$. Inserting these two expressions into Eq.~(\ref{c1c2cond2}), we get $- c'^2_{m,1} (c_{m,2}+c^*_{m,2})/c_{m,1}=\frac{1}{4(N+1)}$, so that $(c_{m,2}+c^*_{m,2})/c_{m,1}<0$. On the other hand, Eq.~(\ref{c1c2cond1}) gives $(c_{m,2}+c^*_{m,2})/c_{m,1}=\frac{1}{4(N+1) c_{m,1}^2}>0$, hence Eqs.~(\ref{c1c2cond1})-(\ref{c1c2cond3}) have no solutions.
\par We thus discard the unphysical vector $(g'_m,h'_m)$ and choose the $N$ physically desired eigenvectors of $M$ as
\begin{eqnarray}\label{gmhm}
(g_k,h_k)&=&c_{k,1}\chi(\alpha_k)+c_{k,2}\xi(\alpha_k),~~~k=1,2,\ldots,N
\end{eqnarray}
where $c_{k,1}$ and $c_{k,2}$ satisfy Eq.~(\ref{c1c2cond1}). It is obvious that $g_k g^\dag_{k'}-h_k h^\dag_{k'}=0$ for $k\neq k'$. In addition, $g_k h^{\mathrm{T}}_{k'}-h_k g^{\mathrm{T}}_{k'}=0$ for any $k$ and $k'$ since $V(\alpha_{k})W^{\mathrm{T}}(\alpha_{k'})-W(\alpha_k)V^{\mathrm{T}}(\alpha_{k'})=0$.
\par In the spirit of Eq.~(\ref{phikMconj2}), it is natural to require that
\begin{eqnarray}
(g_{N+1-k},h_{N+1-k})&=&(g^*_k,h^*_k)=c_{k,1}\chi^*(\alpha_k)+c^*_{k,2}\xi^*(\alpha_k)=c_{k,1}\chi(\alpha_{N+1-k})+c^*_{k,2}\xi(\alpha_{N+1-k}),
\end{eqnarray}
which gives
\begin{eqnarray}
c_{N+1-k,1}=c^*_{k,1}=c_{k,1},~c_{N+1-k,2}=c^*_{k,2},
\end{eqnarray}
and hence
\begin{eqnarray}\label{etalong}
\eta_k&=&\sum^N_{j=1}\{[c_{k,1}V_j(\alpha_k)+c_{k,2}W_j(\alpha_k)]a_j+[c_{k,1}V_j(\alpha_k)-c_{k,2}W_j(\alpha_k)]a^\dag_j\},\nonumber\\
\eta_{N+1-k}&=&\sum^N_{j=1}\{[c_{k,1}V^*_j(\alpha_k)+c^*_{k,2}W^*_j(\alpha_k)]a_j+[c_{k,1}V^*_j(\alpha_k)-c^*_{k,2}W^*_j(\alpha_k)]a^\dag_j\},
\end{eqnarray}
where $k=1,2,\ldots,N/2$ [$k=1,2,\ldots,(N-1)/2$] for even (odd) $N$. The eigenvector for the special mode $k=(N+1)/2$ in the case of odd $N$ is $(g_{(N+1)/2},h_{(N+1)/2})=c_{(N+1)/2,1}\chi(0)+c_{(N+1)/2,2}\xi(0)$.
\subsection{Restoration of the original Hamiltonian $H_B$}
We expect that $H_B$ can be diagonalized as
\begin{eqnarray}
H_D&=&\sum^{N/2~\mathrm{or}~(N-1)/2}_{k=1}\Lambda(\alpha_k)(\eta^\dag_k \eta_k-\eta^\dag_{N+1-k} \eta_{N+1-k}),
\end{eqnarray}
where $\eta_k$ and $\eta_{N+1-k}$ are given by Eq.~(\ref{etalong}). A straightforward calculation leads to
\begin{eqnarray}
&& \eta^\dag_k \eta_k-\eta^\dag_{N+1-k} \eta_{N+1-k} \nonumber\\
&=&\sum_{ij}2i\Im (g^*_{ki}g_{kj}+h_{ki}h^*_{kj}) a^\dag_i a_j+\left[\sum_{ij} i\Im(h^*_{ki}g_{kj}+g_{ki}h^*_{kj})a_ia_j+\mathrm{H.c.}\right],
\end{eqnarray}
where
\begin{eqnarray}
2i\Im (g^*_{ki}g_{kj}+h_{ki}h^*_{kj})&=&2c_{k,1}c^*_{k,2}[W^*_{i}(\alpha_k)V_j(\alpha_k)-V_i(\alpha_k)W^*_j(\alpha_k)]-\mathrm{c.c.}\nonumber\\
&=&8c_{k,1}i^{i-j}[c^*_{k,2}-(-1)^{i-j}c_{k,2}]\sin\frac{\pi ki}{N+1}\sin\frac{\pi kj}{N+1}[e^{\beta(i-j)}-(-1)^{i-j}e^{-\beta(i-j)}],
\end{eqnarray}
and
\begin{eqnarray}
i\Im (h^*_{ki}g_{kj}+g_{ki}h^*_{kj})&=&-c_{k,1}c^*_{k,2}[W^*_{i}(\alpha_k)V_j(\alpha_k)+V_i(\alpha_k)W^*_j(\alpha_k)]-\mathrm{c.c.}\nonumber\\
&=&-4c_{k,1}i^{i-j}[c^*_{k,2}-(-1)^{i-j}c_{k,2}]\sin\frac{\pi ki}{N+1}\sin\frac{\pi kj}{N+1}[e^{\beta(i-j)}+(-1)^{i-j}e^{-\beta(i-j)}].
\end{eqnarray}
\par To recover the OBKC with nearest-neighbor hopping and pairing, we require that the above two quantities both vanish when $i-j$ is even, which can be simply achieved by choosing $c_{k,2}$ to be real for every $k$. For real $c_{k,2}$ and odd $i-j$, we have
\begin{eqnarray}
2i\Im (g^*_{ki}g_{kj}+h_{ki}h^*_{kj})&=&\frac{4}{N+1}i^{i-j} \sin\frac{\pi ki}{N+1}\sin\frac{\pi kj}{N+1}\cosh\beta(i-j),\nonumber\\
i\Im (h^*_{ki}g_{kj}+g_{ki}h^*_{kj})&=&-\frac{2}{N+1} i^{i-j} \sin\frac{\pi ki}{N+1}\sin\frac{\pi kj}{N+1}\sinh\beta(i-j),
\end{eqnarray}
where the condition (\ref{c1c2cond1}) is used. It is interesting to note that any pair of $c_{k,1}$ and $c_{k,2}$ satisfying Eq.~(\ref{c1c2cond1}) results in the same diagonal form $\eta^\dag_k \eta_k-\eta^\dag_{N+1-k} \eta_{N+1-k}$, even though $\eta_k$ and $\eta_{N+1-k}$ both depend on the two.
\par Actually, we can parameterize the two real parameters $c_{k,1}$ and $c_{k,2}$ as
\begin{eqnarray}\label{c1c2para}
c_{k,1}=\frac{1}{2\rho_k\sqrt{2(N+1)}},~~c_{k,2}=\frac{\rho_k}{2\sqrt{2(N+1)}},
\end{eqnarray}
where $\rho_k$ is an arbitrary nonzero real parameter. The explicit form of $\eta_k$ given by Eq.~(\ref{etalong}) then reads
\begin{eqnarray}\label{etaklong}
\eta_k&=&\sum^N_{j=1}\frac{1}{\sqrt{2(N+1)}}(-i)^{j-1}\sin\frac{\pi kj}{N+1}[(\rho^{-1}_k e^{-\beta j}+\rho_k e^{\beta j})a_j+(\rho^{-1}_k e^{-\beta j}-\rho_k e^{\beta j})a^\dag_j]\nonumber\\
&=&\frac{i}{2}\sqrt{\frac{2}{N+1}}\sum^N_{j=1}(-i)^{j}\sin\frac{\pi kj}{N+1}\left\{\left[\rho_k\left(\frac{\delta_+}{ \delta_- } \right)^{j/2}+\frac{1}{\rho_k}\left(\frac{ \delta_- }{\delta_+} \right)^{j/2}\right]a_j-\left[\rho_k\left(\frac{\delta_+}{ \delta_- } \right)^{j/2}-\frac{1}{\rho_k}\left(\frac{ \delta_- }{\delta_+} \right)^{j/2}\right]a^\dag_j \right\}.
\end{eqnarray}
By comparing Eq.~(\ref{etaklong}) with Eq.~(\ref{bnlong}), it can be seen that the $\tilde{b}_k$ in Eq.~(\ref{bnlong}) can be recovered as $\tilde{b}_k=-i\eta_k$ by choosing $\rho_k=\lambda$, which reveals the correspondence between the global free parameter $\lambda$ in the squeezing transformation method and the $k$-dependent free parameter $\rho_k$ in our plane-wave ansatz solution. It should be emphasized that pairing the two modes $k$ and $N+1-k$ into a group provides more degrees of freedom for us in the construction of the quasi-particle operator $\eta_k$, while only one parameter $\lambda$ appears in the squeezing transformation method.
\par Now, using the following relation holding for odd $i-j$,
\begin{eqnarray}
\frac{1}{N+1}\sum^{N/2~\mathrm{or}~(N-1)/2}_{k=1}\sin\frac{\pi k i}{N+1}\sin\frac{\pi k j}{N+1}\sin\frac{\pi k}{N+1}=\frac{1}{8}(\delta_{i-j, 1}+\delta_{i-j, -1}),
\end{eqnarray}
it is easy to show that (for odd $i-j$)
\begin{eqnarray}
\sum^{N/2~\mathrm{or}~(N-1)/2}_{k=1} \Lambda(\alpha_k) 2i\Im (g^*_{ki}g_{kj}+h_{ki}h^*_{kj})&=& \frac{it}{2}(\delta_{i-j, 1}-\delta_{i-j, -1}) ,\nonumber\\
\sum^{N/2~\mathrm{or}~(N-1)/2}_{k=1} \Lambda(\alpha_k) i\Im (h^*_{ki}g_{kj}+g_{ki}h^*_{kj})&=&-\frac{i\Delta}{4}(\delta_{i-j, 1} +\delta_{i-j, -1} ),
\end{eqnarray}
which immediately gives
\begin{eqnarray}
H_D&=&\sum_{ij}\frac{it}{2}(\delta_{i-j, 1}-\delta_{i-j, -1})  a^\dag_i a_j+\left[-\sum_{ij} \frac{i\Delta}{4}(\delta_{i-j, 1} +\delta_{i-j, -1} )a_ia_j+\mathrm{H.c.}\right].
\end{eqnarray}
This completes the proof of $H_D=H_B$.
\section{Conclusions and discussions}\label{SecIV}
\par We have presented an alternative algebraic solution of the open-boundary bosonic Kitaev chain using a combination of the standard Bogoliubov transformation and an asymmetric plane-wave ansatz. The eigenvalue problem of the non-Hermitian associated matrix is solved analytically, yielding closed-form expressions for the eigenvalues and eigenvectors. We have resolved the twofold degeneracy of each eigenvalue by imposing the bosonic commutation relations, proving that only one physical Bogoliubov mode can be constructed. The resulting Bogoliubov transformation exhibits non-uniquenesses reflected in the $k$-dependent parameterization given by Eq.~(\ref{c1c2para}), which in the homogeneous case corresponds exactly to the freedom in the local squeezing transformations used in Ref.~\cite{PRX2018}. Our method works directly in Fock space for any finite chain length $N$, complementing both the squeezing transformation (followed by gauge and Fourier transformations) and the non-Bloch band theory.
\par We emphasize that the asymmetric plane-wave ansatz approach proposed in this work is not limited to the OBKC, but may be applied to more general bosonic pairing models. Moreover, in a series of recent studies of two-magnon problems in various higher-spin spin chains~\cite{PRB2024_1,PRB2024_2,physica2025}, the symmetric plane-wave ansatz has been recently employed to solve a class of inhomogeneous Bloch Hamiltonians. To illustrate the power of our method in the inhomogeneous case, let us consider a minimal model with the hopping and pairing strength on the first bond different from the others,
\begin{eqnarray}\label{Hform1_inh}
H'_{B}&=&  \frac{it'}{2}(a^\dag_2 a_1-a^\dag_1 a_2)+\frac{1}{2}\sum^N_{i,j=2}\frac{i\Delta'}{2}[( a^\dag_1 a^\dag_2 - a_2 a_1)+( a^\dag_2 a^\dag_1 - a_1 a_2)]\nonumber\\
&&+\sum^N_{i,j=2}\frac{it}{2}(\delta_{i-j,1}-\delta_{j-i,1})a^\dag_i a_j+\frac{1}{2}\sum^N_{i,j=2}\frac{i\Delta}{2}(\delta_{i-j,1}+\delta_{j-i,1})( a^\dag_i a^\dag_j - a_j a_i),
\end{eqnarray}
where $t'\neq t$ and $\Delta'\neq\Delta$. In the framework of the asymmetric plane-wave ansatz, the resulting bulk and boundary equations for $F_1$ become
\begin{eqnarray}\label{ }
V_j\Lambda=i(-\delta_-V_{j-1}+\delta_+V_{j+1}),~j=3,\ldots,N-1
\end{eqnarray}
and
\begin{eqnarray}\label{ }
V_1\Lambda&=&i\delta'_+ V_2,\nonumber\\
V_2\Lambda&=&i(-\delta'_-V_{1}+\delta_+V_{3}),\nonumber\\
V_N\Lambda&=&-i\delta_- V_{N-1}.
\end{eqnarray}
A similar and straightforward analysis shows that we can extend the homogenous wave function (\ref{VjXm}) to the following form,
\begin{eqnarray}\label{VjXm_inh}
V_j(q)=2(-i)^{j-1}\sin(qj+f_q)e^{-\beta j},
\end{eqnarray}
where $\beta$ is still determined by Eq.~(\ref{expB}) and $f_q$ is a newly introduced scattering phase shift due to the boundary defect on the first bond. The wave number $q$ is obtained by inserting Eq.~(\ref{VjXm_inh}) into the bulk and boundary equations. The $N-3$ bulk equations simply give the eigenvalue $\Lambda(q)=2\sqrt{\delta_+\delta_-}\cos q$. By eliminating $V_1$ from the first two boundary equations and using the ansatz, we get
\begin{eqnarray}\label{ }
\sin (2q+f_q) \left(4\cos^2 q -1-\frac{\delta'_- \delta'_+}{\delta_+\delta_-}\right)&=& \sin (4q+f_q),\label{qfq}\\
\sin [q(N+1)+f_q]&=& 0.\label{qfq1}
\end{eqnarray}
From Eq~(\ref{qfq1}) we have
\begin{eqnarray}\label{ }
q=\frac{\pi m-f_q}{N+1},~m\in\mathbb{Z}
\end{eqnarray}
showing that $f_q$ indeed plays the role of a phase shift. By inserting $f_q=m\pi-(N+1)q$ into Eq.~(\ref{qfq}), we obtain the quantization condition determining $q$,
\begin{eqnarray}\label{qcond}
\sin [q(N+1)]+\sin [(N-1)q] \left(1-\frac{\delta'_- \delta'_+}{\delta_+\delta_-}\right)&=& 0.
\end{eqnarray}
\par Considering this, we expect the asymmetric plane-wave ansatz can also be used to solve bosonic pairing models with more general inhomogeneous structures near the chain ends (e.g., with finite boundary on-site energies), where the application of the squeezing transformation method or the non-Bloch band theory may become indirect due to the lack of periodicity in the bulk. We will leave these studies to future works.\\
\section*{Acknowledgements}
This work was supported by the National Key Research and Development Program of China under Grant No. 2021YFA1400803 and by the Innovation Program for Quantum Science and Technology under Grant No. 2023ZD0300700.
\appendix
\section{Proof of Eqs.~(\ref{VW1}) and (\ref{VW2})}\label{AppA}
From Eqs.~(\ref{VjXm}) and (\ref{WjXm}), we have
\begin{eqnarray}
&&V(\alpha_m)W^\dag(\alpha_n)\nonumber\\
&=&2\sum_{j}[\cos(\alpha_m-\alpha_n)j-(-1)^j\cos(\alpha_m+\alpha_n)j ]\nonumber\\
&=&2\frac{\cos\frac{(N+1)(\alpha_m-\alpha_n)}{2}\sin\frac{N(\alpha_m-\alpha_n)}{2}}{\sin\frac{(\alpha_m-\alpha_n)}{2}} +1+(-1)^{N+1}\left[\cos (N+1)(\alpha_m+\alpha_n)+\sin (N+1)(\alpha_m+\alpha_n) \tan\frac{\alpha_m+\alpha_n}{2}\right].\nonumber
\end{eqnarray}
\par From $ \alpha_m-\alpha_n=(m-n)\pi/(N+1)$, we have
\begin{eqnarray}
2\frac{\cos\frac{(N+1)(\alpha_m-\alpha_n)}{2}\sin\frac{N(\alpha_m-\alpha_n)}{2}}{\sin\frac{(\alpha_m-\alpha_n)}{2}}= \begin{cases}
-[(-1)^{m-n}+1], & m\neq n,\\
2N, & m=n.
  \end{cases}
\end{eqnarray}
\par From $(N+1)(\alpha_m+\alpha_n)=[m+n-(N+1)]\pi$, we have
\begin{eqnarray}
\cos (N+1)(\alpha_m+\alpha_n) &=&(-1)^{N+1}(-1)^{m-n},\nonumber\\
\sin (N+1)(\alpha_m+\alpha_n) &=&0.
\end{eqnarray}
\par Combining these equations, we get
\begin{eqnarray}
V(\alpha_m)W^\dag(\alpha_n)&=&  \begin{cases}
0, & m\neq n,\\
2(N+1), & m=n.
  \end{cases}
\end{eqnarray}
\par By observing that $W^{\mathrm{T}}(\alpha_n)=W^{*\dag}(\alpha_n)=W^{\dag}(-\alpha_n)$, we immediately get Eq.~(\ref{VW2}).

\end{document}